\documentclass[prd,preprint, superscriptaddress,preprintnumbers,nofootinbib]{revtex4}
\usepackage{graphicx}
\usepackage{epsfig}
\usepackage{bm}
\usepackage{latexsym,amssymb,amsmath,amssymb,wasysym,float}

\usepackage{color}

\usepackage{enumitem}

\newcommand{\postscript}[2]{\setlength{\epsfxsize}{#2\hsize}
   \centerline{\epsfbox{#1}}}

\usepackage[usenames,dvipsnames]{xcolor}
\definecolor{orange}{cmyk}{0,0.5,1,0}
\definecolor{rossoCP3}{cmyk}{0,.88,.77,.40}
\definecolor{graa}{rgb}{0.8,0.8,0.8}
\definecolor{blaa}{rgb}{0.2,0.2,0.6}

\begin{document}

\preprint{MPP-2024-5}
\preprint{LMU-ASC 01/24}

\title{\color{rossoCP3} The Dark Dimension, the Swampland, and the
  Dark Matter Fraction Composed of Primordial Near-Extremal Black Holes}

\author{\bf Luis A. Anchordoqui}

\affiliation{Department of Physics and Astronomy,\\  Lehman College, City University of
  New York, NY 10468, USA
}


\affiliation{Department of Astrophysics,
 American Museum of Natural History, NY
 10024, USA
}

\author{\bf Ignatios Antoniadis}

\affiliation{High Energy Physics Research Unit, Faculty of Science, Chulalongkorn University, Bangkok 1030, Thailand}

\affiliation{Laboratoire de Physique Th\'eorique et Hautes \'Energies
  - LPTHE \\
Sorbonne Universit\'e, CNRS, 4 Place Jussieu, 75005 Paris, France
}
\affiliation{Center for Cosmology and Particle Physics, Department
  of Physics,\\ New York University, 726 Broadway, New York, NY 10003, USA}

\author{\bf Dieter\nolinebreak~L\"ust}

\affiliation{Max--Planck--Institut f\"ur Physik,  
 Werner--Heisenberg--Institut,
80805 M\"unchen, Germany
}

\affiliation{Arnold Sommerfeld Center for Theoretical Physics, 
Ludwig-Maximilians-Universit\"at M\"unchen,
80333 M\"unchen, Germany
}

\begin{abstract}
  \vskip 2mm \noindent In a recent publication we studied the decay rate of
  primordial black holes perceiving the dark dimension, an innovative
  five-dimensional (5D) scenario that has a compact space with characteristic length-scale in the micron range. We demonstrated
that the rate of Hawking radiation of 5D black holes
slows down compared to 4D black holes of the same
mass. Armed with our findings we showed 
that for a species scale of ${\cal O} (10^{10}~{\rm GeV})$, an
all-dark-matter interpretation in terms of primordial black holes
should be feasible for black hole masses in the range $10^{14} \lesssim M/{\rm
  g} \lesssim 10^{21}$. As a natural outgrowth of our recent study,
herein we calculate the Hawking evaporation of near-extremal 5D black
holes. Using generic entropy arguments  we demonstrate that Hawking
evaporation of higher-dimensional near-extremal black holes proceeds
at a slower rate than the corresponding Schwarzschild black holes of
the same mass. Assisted by this result we show that if there were 5D primordial
near-extremal black holes in nature, then a primordial black hole all-dark-matter
interpretation would be possible in the mass range  $10^{5}
\sqrt{\beta} \lesssim M/{\rm
  g} \lesssim 10^{21}$, where $\beta$ is a parameter that controls the difference between mass and charge of the associated near-extremal black hole.

\end{abstract}

\maketitle 

\section{Introduction}

The Swampland program aims at
understanding which are the ``good''
low-energy efective field theories (EFTs) that can couple to gravity consistently (e.g. the
landscape of superstring theory vacua) and distinguish them from the
``bad'' ones that cannot~\cite{Vafa:2005ui}. In theory space, the
boundary setting apart the good theories from those downgraded to the
swampland is characterized by a set of conjectures classifying the
indispensable properties of an  
 EFT to enable a consistent
completion into quantum gravity. These conjectures provide a catwalk from quantum gravity to astrophysics, cosmology, and particle
physics~\cite{Palti:2019pca,vanBeest:2021lhn,Agmon:2022thq}.

For instance, the distance conjecture (DC) predicts the appearance of infinite towers of states that become
exponentially light and trigger the collapse of the EFT at infinite distance limits
in moduli space~\cite{Ooguri:2006in}. Associated to the DC is the anti-de
Sitter (AdS) distance conjecture, which
correlates the dark energy density to the mass scale $m$ characterizing the infinite tower of states,
$m \sim |\Lambda|^\alpha$, as the negative AdS vacuum energy
 $\Lambda \to 0$, with $\alpha$ a positive constant of ${\cal O} 
 (1)$~\cite{Lust:2019zwm}. In addition, under the premise that
 this scaling behavior holds in de Sitter (dS)  -- or
 quasi dS -- space, an unbounded number of massless modes also materialize in the limit $\Lambda \to 0$.

As demonstrated in~\cite{Montero:2022prj}, applying the AdS-DC to dS space could help
elucidate the origin of the cosmological hierarchy $\Lambda/M_p^{4} \sim 10^{-120}$, because it connects the
size of the compact space $R_\perp$ to the
dark energy scale $\Lambda^{-1/4}$ via
$R_\perp \sim \lambda \ \Lambda^{-1/4}$,
where the proportionality factor is estimated to be within the range
$10^{-1} < \lambda < 10^{-4}$. Actually, the previous relation between
$R_\perp$ and $\Lambda$ derives from
  constraints by theory and experiment. On the one hand, since the associated Kaluza-Klein (KK) 
  tower contains massive spin-2 bosons, the Higuchi
  bound~\cite{Higuchi:1986py}  provides an absolute upper limit to
  $\alpha$, whereas explicit string calculations
of the vacuum energy~(see
e.g.~\cite{Itoyama:1986ei,Itoyama:1987rc,Antoniadis:1991kh,Bonnefoy:2018tcp})
yield a lower bound on $\alpha$. All in all, the theoretical
constraints lead to $1/4 \leq \alpha \leq 1/2$. On the other hand,
experimental arguments (e.g. constraints on deviations from Newton's
gravitational inverse-square law~\cite{Lee:2020zjt} and neutron star
heating~\cite{Hannestad:2003yd}) lead to the conclusion encapsulated
in $R_\perp \sim \lambda \ \Lambda^{-1/4}$; namely, that there is one extra dimension of
radius $R_\perp$ is in the micron range, and that the lower bound for $\alpha =
1/4$ is basically saturated~\cite{Montero:2022prj}. A theoretical
amendment on the connection between the cosmological and KK mass scales confirms $\alpha =
1/4$~\cite{Anchordoqui:2023laz}. Assembling all this together, we can conclude that the KK tower of the new (dark)
dimension opens up at the mass scale $m_{\rm KK} \sim
1/R_\perp$. For the dark dimensions scenario, the five-dimensional (5D) Planck scale (or species scale
where gravity becomes strong~\cite{Dvali:2007hz,Dvali:2007wp,Cribiori:2022nke,vandeHeisteeg:2023dlw}) is
given by
\begin{equation}
M_* \sim m_{\rm KK}^{1/3} \ M_p^{2/3} \, ,
\end{equation}
where $M_p$ is the reduced Planck mass. Thus, since the size of dark dimension in the micron scale, $m_{\rm
  KK} \sim 1~{\rm eV}$ and so $10^9 \alt M_*/{\rm GeV} \alt  10^{10}$.

Early universe phenomena and the nature of dark matter are among the
most strategic science cases of theoretical high energy physics. As
potentially the first density perturbations to collapse during the
early universe, primordial black holes (PBHs) provide our earliest
landmarks to probe the very early universe, at energies between the QCD
phase transition and the Planck scale. The corresponding energy scales
are out of the reach of existing cosmological probes. Much of the parameter space characterizing the PBH abundance
has been constrained by existing probes, but a large window remains
open, where PBHs around asteroid mass ($10^{-15}$ to $10^{-10}
M_\odot$) could make up the entirety of dark matter~\cite{Carr:2020xqk,Green:2020jor,
  Villanueva-Domingo:2021spv,LISACosmologyWorkingGroup:2023njw}. The detection of
PBHs could provide a cornerstone for our perception of
the physics processes in the very early universe. This significant
reward motivates new investigations on this subject.

In previous work~\cite{Anchordoqui:2022txe,Anchordoqui:2022tgp}, we
first calculated the decay rate of
PBHs perceiving the dark dimension and demonstrated
that the rate of Hawking radiation slows down compared to 4D black holes of the same
mass. Then, we used this result to show that the mass range supporting a 5D PBH
all-dark-matter interpretation is extended compared to that in the 4D
theory by 3 orders of magnitude in the low mass region. As a natural
outgrowth of this work, herein we study the Hawking evaporation of near-extremal 5D black
holes.  More concretely, we generalize the 4D results obtained
in~\cite{deFreitasPacheco:2023hpb}  to $d$-dimensions. We then discuss
the impact of our findings in assessing the dark matter fraction that
could be composed of PBHs. Since Hawking evaporation
of near-extremal 5D black holes proceeds at a slower rate
than the corresponding Schwarzschild black holes of the same mass, we
show herein that near extremality could further relax the lower mass bound range of
a PBH all-dark-matter interpretation.

The outline of the paper is as follows. In Sec.~\ref{sec:2} we
summarized the results of our previous work. In Sec.~\ref{sec:3} we
provide an overview of near-extremal black
holes and discuss the various charges that can potentially bring 
together the
inner and outer horizons. In Sec.~\ref{sec:4} we lay out a proof of principle
for primordial near-extremal black holes investigating a model in
which the charge leading to extremality is carried by dark electrons
living in the bulk. In Sec.~\ref{sec:5}, we first adopt generic
entropy arguments to derive the scaling behavior of the decay rate of
higher-dimensional near-extremal
black holes. After that, armed with our findings we investigate how near-extremal black
holes perceiving the dark dimension could modify the constraints on
a PBH all-dark-matter interpretation. The paper wraps up in Sec.~\ref{sec:6} with some conclusions.

\section{Primordial black hole dark matter interpretation}

\label{sec:2}

It has long been speculated that black holes could be produced from the
collapse of large amplitude fluctuations in the early
universe~\cite{Zeldovich:1967lct,Hawking:1971ei,Carr:1974nx,Carr:1975qj}. For
an order of magnitude estimate of the black hole mass $M$, we first
note that the cosmological energy density scales with time $t$
as $\rho \sim 1/(Gt^2)$ and the density  needed for a region of mass
$M$ to collapse within its Schwarzschild radius is $\rho \sim
c^6/(G^3M^2)$, so that PBHs would initially
have around the cosmological horizon mass~\cite{Carr:2020xqk}
\begin{equation}
M \sim \frac{c^3 t}{G} \sim 10^{15}
\left(\frac{t}{10^{-23}~{\rm s}}\right)~{\rm g} \, ,
\end{equation}
with $M_p = 1/\sqrt{8 \pi G}$. This means that a black hole would have the Planck mass ($M_p
\sim 10^{-5}~{\rm g}$) if they formed at the Planck time
($10^{-43}~{\rm s}$), $1~M_\odot$ if they formed at the QCD epoch
($10^{-5}~{\rm s}$), and $10^{5} M_\odot$ if they formed at $t \sim
1~{\rm s}$, comparable to the mass of the holes thought to reside in galactic nuclei. This back-of-the-envelope calculation suggests that PBHs could span an enormous mass
range. Despite the fact that the mass spectrum of these PBHs is yet to
be shaped, on cosmological scales they would behave like a typical cold dark
matter particle.

Nevertheless, an all-dark-matter interpretation in terms of
PBHs is severely constrained by
observations~\cite{Carr:2020xqk,Green:2020jor,
  Villanueva-Domingo:2021spv,LISACosmologyWorkingGroup:2023njw}. Of relevance to our investigation, the extragalactic
$\gamma$-ray background~\cite{Carr:2009jm} and  the 
spectrum of the cosmic microwave background (CMB)~\cite{Clark:2016nst} constrain
PBH evaporation of black holes with masses $\lesssim 10^{17}~{\rm g}$,
whereas the non-observation of microlensing events from the MACHO~\cite{Macho:2000nvd}, EROS~\cite{EROS-2:2006ryy},
  Kepler~\cite{Griest:2013aaa}, Icarus~\cite{Oguri:2017ock}, OGLE~\cite{Niikura:2019kqi} and Subaru-HSC~\cite{Croon:2020ouk} collaborations constrain black holes with masses $\gtrsim
10^{21}~{\rm g}$. 

Microscopic black holes of
Schwarzschild radii smaller than the size of the dark dimension are: bigger, colder, and longer-lived than a
usual 4D black hole of the same
mass~\cite{Argyres:1998qn}. Indeed, Schwarzschild black holes radiate all particle species lighter than or comparable to
their temperature, which in four dimensions is related to the mass of the black hole by
\begin{equation} 
T_s = \frac{M_p^2}{8 \pi M} \sim 
\, \bigg(\frac{M}{10^{16} \, {\rm g}} \bigg)^{-1}~{\rm MeV}\,,
\label{T4d}
\end{equation}
whereas for five dimensional black holes the temperature mass relation
is found to be~\cite{Anchordoqui:2022txe}
\begin{equation}
  T_s  \sim \frac{1}{r_s} \sim  
    \left(\frac{M}{10^{12}~{\rm g}}\right)^{-1/2}~{\rm MeV} \, ,
\label{tempes}
  \end{equation}
  where
  \begin{equation}
    r_s(M) \sim \frac{1}{M_*} \left[ \frac{2}{3\, \pi}
   \ \frac{M}{M_*}  \right]^{1/2}\end{equation}  
is the 5D Schwarzschild radius~\cite{Myers:1986un}. The numerical estimate of (\ref{tempes})  applies to
the dark dimension scenario with $M_* \sim 10^{10}~{\rm GeV}$,
which is consistent with 
astrophysical
observations~\cite{Anchordoqui:2022ejw,Noble:2023mfw}.\footnote{We have taken the
highest possible value of $M_*$ to remain conservative in the estimated bound on $f_{\rm PBH}$.}  It is evident that 5D
black holes are colder than 4D black holes of the same mass. The
Hawking radiation causes a 4D black hole to lose mass at the following rate~\cite{Keith:2021guq}
\begin{eqnarray}
 \left. \frac{dM}{dt}\right|_{\rm evap} & = & -\frac{M_p^2}{30720 \ \pi \ M^2} \ \sum_{i} c_i(T_s) \ \tilde f \ \Gamma_s
    \nonumber \\ 
   & \sim & -7.5 \times 10^{-8} \ \left(\frac{M}{10^{16}~{\rm
                                                   g}}\right)^{-2} \ \sum_{i}
      c_i (T_s)  \ \tilde f
 \ \Gamma_s~{\rm g/s}   \,, 
\label{cinco}
\end{eqnarray}
whereas a 5D black hole has an evaporation rate of~\cite{Anchordoqui:2022txe}
\begin{eqnarray}
  \left. \frac{dM}{dt}\right|_{\rm evap}
& \sim & - 9 \ \pi^{5/4} \zeta(4) T_s^2 \ \sum_{i} c_i(T_s) \ \tilde f
      \ \Gamma_s 
   \sim  -\frac{27 \ \Lambda^{1/4} \ M_p^2}{64 \
      \pi^{3/4} \ \lambda \ M} \ \sum_{i} c_i(T_s) \ \tilde f \ \Gamma_s
    \nonumber \\
& \sim & - 1.3 \times 10^{-12} \ \left(\frac{M}{10^{16}~{\rm g}}\right)^{-1} \ \sum_{i}
      c_i (T_s) \ \tilde f
\  \Gamma_s~{\rm g/s}  \,,     
\label{seis}
\end{eqnarray}
where $c_i(T_s)$ counts the number of internal degrees of freedom of particle 
species $i$ of mass $m_i$ satisfying $m_i \ll T_s$,  $\tilde f = 1$  $(\tilde f = 7/8)$ for bosons
(fermions), and where $\Gamma_{s=1/2} \approx 2/3$ and $\Gamma_{s=1} \approx
1/4$ are the (spin-weighted) dimensionless greybody factors normalized to the black
hole surface area~\cite{Anchordoqui:2002cp}. In the spirit
of~\cite{Emparan:2000rs}, we neglect KK graviton emission because the KK modes are excitations in
the full transverse space, and so their overlap with the small
(higher-dimensional) black holes is suppressed by the geometric factor
$(r_s/R_\perp)$ relative to the brane fields. Thus, the geometric
suppression precisely compensates for the enormous number of modes,
and the total contribution of all KK modes is only the same order as
that from a single brane field. On top of that, the 5D graviton has 5 helicities,
but the spin-1 helicities do not have zero modes, because we assume
the compactification has $S^1/\mathbb{Z}_2$ symmetry and so the $\pm
1$ helicities are projected out. The greybody factors of spin-2
particles strongly suppress massless graviton emission on the brane
$\Gamma_{s=2}/\Gamma_{s=1/2} \alt 10^{-3}$, and the emission of $\pm
1$ helicities in the bulk is also suppressed; see e.g., Fig.~2 of
Ref.~\cite{Ireland:2023zrd}. Contribution from the spin-0 depends on the radion mass. Since the
addition of one scalar does not modify the order of magnitude
calculations of this work, throughout we neglect the graviton
emission. At the end of the concluding section we comment on the
feasibility of detecting graviton emission on the brane. Now, comparing (\ref{cinco}) and (\ref{seis}) it is easily seen that 5D black holes live
longer than 4D black holes of the same mass.

Integrating (\ref{seis}) we can parametrize the 5D black hole
lifetime as a function of its mass and temperature,
\begin{equation}
  \tau_s \sim 13.8 \ \Bigg(\frac{M}{10^{12}~{\rm g}} \Bigg)^2 \ \left(\frac{6}{\sum_{i}
      c_i (T_s) \ \tilde f
      \  \Gamma_s}\right)~{\rm Gyr} \, ,
\label{lifetime}  
\end{equation}
where we have used (\ref{tempes}) to estimate that
$T_ s\sim 1~{\rm MeV}$ and therefore $c_i(T_s)$ receives a
contribution of 6 from neutrinos, 4 for electrons, and 2 from photons,
yielding $\sum_{i} c_i (T_s) \ \tilde f \ \Gamma_s = 6$.  Armed with
(\ref{lifetime}) we can estimate the bound on the 5D PBH abundance by
a simple rescaling procedure of the $d=4$ bounds on the fraction of
dark matter composed of primordial black holes $f_{\rm PBH}$. The key
point for such a rescaling is that for a given photon energy, or
equivalently a given Hawking temperature, we expect a comparable limit
on $f_{\rm PBH}$ for both $d=4$ and $d=5$. For example, from
(\ref{T4d}) and (\ref{tempes}) we see that the constraint of
$f_{\rm PBH} \alt 10^{-3}$ for 4D black holes with
$M_{\rm BH} \sim 10^{16}~{\rm g}$, should be roughly the same for the
abundance of 5D black holes with $M_{\rm BH} \sim 10^{12}~{\rm
  g}$. Now, since in $d=4$ for $M_{\rm BH} \sim 10^{17}~{\rm g}$ we
have $f_{\rm PBH} \sim 1$, this implies the same abundance for 5D
black holes of $M_{\rm BH} \sim 10^{14}~{\rm g}$. By duplicating this
procedure for heavier black holes we conclude that for a species scale
of ${\cal O} (10^{10}~{\rm GeV})$, an all-dark-matter interpretation in
terms of 5D black holes must be feasible for masses in the range
\begin{equation}
10^{14} \lesssim M/{\rm
  g} \lesssim 10^{21} \, .
\label{massrange}
\end{equation}
This range is extended compared to that in the
4D theory by 3 orders of magnitude in the low mass region.

At this stage, it is worthwhile to point out that a stunning coincidence is that the size of the dark dimension
  $R_\perp \sim$ wavelength of visible light. This means that the
  Schwarzschild radius of 5D black holes is well below the wavelength
  of light. For point-like lenses, this is the critical length where
  geometric optics breaks down and the effects of wave optics suppress
  the magnification, obstructing the sensitivity to 5D PBH
  microlensing signals~\cite{Croon:2020ouk}.

\section{Near-extremal black holes}
\label{sec:3}

Asymptotically flat, static, and spherically symmetric charged (or rotating) black
holes can be categorized as generalizations of the popular
Schwarzschild metric. Such charged black holes carry additional quantum
numbers, which make their properties change drastically and unique new
phenomena arise. A far reaching hallmark of rotating black holes or those which are electrically
(and/or magnetically) charged is their thermodynamical property dubbed
extremality (i.e. zero temperature). Extremal black holes are in essence stable gravitational objects with finite entropy but
vanishing temperature, and so the contribution to the
gravitational energy completely originates in the electromagnetic
charges and/or rotational angular momentum/spin.\footnote{Actually, when the temperature reduces to zero, the entropy reduces to the logarithm of the number of degenerate ground states, which is zero if the ground state is not degenerate~\cite{Hawking:1994ii,Page:2000dk}.} Extremality also
implies that the inner (Cauchy) and outer (event) horizons do
coincide, leading to a vanishing surface gravity. The Reissner-Nordstr\"om (RN) metric describes the simplest extremal black hole, which has its mass equal to its charge in appropriate units.

It has long been suspected that any electromagnetic charge or
spin would be lost very quickly by any 4D black hole
population of 
primordial origin. On the one hand, the electromagnetic charge of a
Reissner-Nordstrom (RN) black hole is spoiled by the Schwinger
effect~\cite{Schwinger:1951nm}, which allows pair-production of
electron-positron pairs in the strong electric field outside the
black hole, leading to the discharge of the black hole and subsequent
evaporation~\cite{Gibbons:1975kk,Hiscock:1990ex}. On the other hand, a rapidly
rotating Kerr black hole~\cite{Kerr:1963ud} spins down to a nearly non-rotating state
before most of its mass has been given up, and therefore it does not
approach to extremal when it evaporates~\cite{Page:1976ki}. All in
all, near-extremal primordial RN black holes or Kerr black holes are
not expected to prevail in the universe we live in.

Adding to the story, it was pointed out in~\cite{Alonso-Monsalve:2023brx} that primordial
black holes could grow by absorbing unconfined quarks and
gluons. Given Debye screening, the quark-gluon plasma must be color
neutral on long length scales $l \gg \lambda_D$, but could have a nontrivial distribution of color charge across shorter length scales
$l \sim \lambda_D$. In particular, there could  exist regions with net
color charge, whose spatial extent is set by $\lambda_D (T
)$~\cite{Manuel:2003zr,Manuel:2004gk}. If this were the case, then
black holes would acquire a net color
charge~\cite{Alonso-Monsalve:2023brx}. However, after the QCD
confinement transition, the medium would cease to screen the
primordial black hole enclosed charge ($\lambda_D \to \infty$), and
therefore it would become energetically (very) costly for any
primordial black hole to maintain its color charge.

An alternative
interesting possibility is to envision a scenario where the black hole
is charged under a generic unbroken $U(1)$ symmetry (dark photon),
whose carriers (dark electrons with a mass $m'_e$ and a gauge coupling
$e'$) are always much heavier than the temperature of the black
hole~\cite{Bai:2019zcd}. This implies that the charge $Q$ does not get
evaporated away from the black hole and remains therefore
constant. Strictly speaking, the pair production rate per unit volume
from the Schwinger effect can be slowed down by arbitrarily decreasing
$e'$, whereas the weak gravity conjecture (WGC) imposes a constraint on the
charge per unit mass; namely, for each conserved gauge charge there
must be a sufficiently light charge carrier such that
\begin{equation}
e'q/m_{e'} \geq
\sqrt{4\pi} \sqrt{(d-3)/(d-2)} M_p^{-(d-2)/2} \,,
\label{dWGC}
\end{equation}
where $q$ is the
integer-quantized electric charge of the
particle and $M_p$ is the reduced Planck mass~\cite{Arkani-Hamed:2006emk,Harlow:2022ich}. Setting $e = e' = \sqrt{4 \pi
  \alpha}$ the (4D) Schwinger effect together with the WGC lead to a bound on the minimum black hole mass of near
extremal black holes with evaporation time longer than the age of
Universe, $M_{ne}  \gtrsim 5 \times 10^{15}~{\rm g} (m_{e'}/10^9~{\rm
  GeV})^{-2}$~\cite{Bai:2019zcd}.

The latest chapter in the story is courtesy of dS backgrounds. If a black hole is
embedded in a dS background, there is an additional bound on $m_{e'}$
from the festina lente (FL) conjecture~\cite{Montero:2019ekk}. This is
because the RN-dS line element comprises two horizons
accessible to an observer outside the black hole: {\it (i)}~the
familiar event horizon of the charged black hole and {\it (ii)}~the
cosmological horizon. Usually, the black hole and the cosmological
horizons would have different temperatures, and so they cannot be in
thermal equilibrium. Considering large black holes whose size is
comparable to the dS radius and demanding their evaporation  avoids
superextremality leads to the festina lente bound: for every charged
state in the theory,
\begin{equation}
  m_{e'}^4 \gtrsim (e'q)^2  \frac{(d-1) (d-2)}{2 \ M_p^{2-d} \ \ell_d^2}  \,,
\label{festina-lente}
\end{equation}
where $\ell_d$ is the dS radius. This bound is satisfactorily satisfied in our
universe for the electron.

\section{Higher-dimensional Schwinger pair production}
\label{sec:4}

The metric of a $d$-dimensional RN-like dS black hole has the form
\begin{equation}
  ds^2 = U(r) \ dt^2 - U^{-1} (r) \ dr^2
- d\Omega_{d-2}^2 \,,
\label{RS-metric}
\end{equation}
where $d\Omega_{d-2}$ is the line element of a flat space of $d-2$ dimensions in spherical coordinates and
\begin{equation}
  U(r) = 1 - \frac{2M}{M_*^{d-2} \ r^{d-3}} +
  \frac{(e'Q)^2}{4\pi \ M_*^{d-2} \ r^{2d-6}} - \frac{r^2}{\ell_d^2} \, ,
\label{uofr}
\end{equation}
where $M_*$ is the $d$-dimensional Planck scale and the coupling $e'$
is taken as a parameter. A well motivated scenario emerges if the SM
has charges under the $U(1)$ field, such that $e'$ becomes of
the order of SM gauge couplings divided by the square root of the
volume of the internal space.

Before proceeding, we pause to note that the FL inequality (\ref{festina-lente}) remains the
same in any number of dimensions, since the gauge coupling has units
of Energy$^{2-d/2}$~\cite{Montero:2021otb}. However, it is important to stress that
the FL bound only applies to black holes of size comparable to the
cosmological horizon and therefore it is not of direct interest for scales
smaller than $R_\perp$. The dS-WGC is relevant  to black holes with a horizon radius smaller than
$R_\perp$~\cite{Antoniadis:2020xso}. However, for large $\ell_4$ values,
dS-WGC constraints on the particle spectrum can also be safely neglected. Hereafter, we
proceed under the assumption of a
(nearly)  flat 5D Minkowski background and neglect the
last term in (\ref{uofr}). We further assume that (\ref{dWGC}) is
satisfied. For details, see the Appendix.

For $d$ dimensions, the Schwinger probability (per unit volume and unit
time) of pair creation in a constant electric field is found to be
\begin{equation}
\Gamma_d  = \frac{(2J+1)}{(2\pi)^{d-1}} \ \sum_{n=1}^\infty
(-1)^{(2J+1) (n+1)} \
  \left(\frac{e' E'}{n}\right)^{d/2} \ \exp\left( -\frac{\pi n m_{e'}^2}{e'E'} \right)\,,
\label{Gammad}
\end{equation}
where $E'$ is the dark electric field, $e'Q \to \sqrt{4 \pi}
M /M_*^{(d-2)/2}$, and $J$ is the spin of the produced
particles~\cite{Bachas:1992bh}. Throughout, the arrow indicates we are considering
near-extremal rather than extremal black holes. A point
worth noting at this juncture is that for $d>6$ the RN-dS solution is
gravitationally unstable~\cite{Konoplya:2008au,Konoplya:2013sba} and so
we focus our calculation on the interesting case of $d=5$ that characterizes the dark dimension
scenario~\cite{Montero:2022prj}. For $d=5$, the spin $J$ is half-integer and so (\ref{Gammad}) can be rewritten as
\begin{equation}
  \Gamma_5 = \frac{1}{8 \pi^4}  \sum_{n=1}^\infty \ 
  \left(\frac{e' E'}{n}\right)^{5/2} \ \exp\left( -\frac{\pi n
      m_{e'}^2}{e'E'} \right) \, .
\label{G5}
\end{equation}
It is of interest to make a comparison between the outer horizon radius of the 5D black hole,
\begin{equation}
r_{+,5d} = \left[\frac{M + \sqrt{M^2 - (e'Q)^2 \ M_*^3/(4\pi)}}{M_*^3}\right]^{1/2} \to
  \sqrt{\frac{M}{M_*^3}} \, ,
\end{equation}
and that of a 4D black hole $r_{+,4d} \to M/M_p^2$ with
the same $M$ and $e'Q$. It follows that $r_{+,5d} > r_{+,4d} \Leftrightarrow M < M_p^4/M_*^3$, which if we take $M_* \sim 10^9~{\rm GeV}$
implies that $M < 10^{45}~{\rm GeV}$ and $r_{+,5d} < 1~\mu{\rm
  m}$. This in turn entails that for the length scale of interest, the
 outer horizon of a 5D RN black hole is larger than the
corresponding 4D black hole. If this were the case, then the electric
field strength in the outer horizon would be smaller and it would be easier
to suppress Schwinger production pairs in five than in four
dimensions.

\section{Higher-dimensional near-extremal black hole decay rate}
\label{sec:5}

The suppression of the near-extremal black hole decay rate with
respect to that of Schwarzschild black holes of the same mass
advertised in the Introduction is evident in the order of magnitude calculation that follows.

For a $d$-dimensional spacetime, the relation between the black hole entropy $S$ and its mass $M$ is~\cite{Anchordoqui:2001cg}
\begin{equation}
  S = 4 \pi \ M \ r_s/ (d-2) \sim (M/M_*)^{(d-2)/(d-3)} \, .
\end{equation}
For a Schwarzschild black hole, the temperature scales with entropy as
\begin{equation}
  T_s \sim M_* \ S^{-1/(d-2)}
\end{equation}
and the black hole decay rate scales as
\begin{equation}
  \Gamma_s \sim T_s \, .
\end{equation}
For near-extremal black holes, however, the temperature scales as
\begin{equation}
  T_{ne} \sim \frac{c}{S}
\end{equation}
where $c = \sqrt{M^2 - (e'Q)^2 M_*^{d-2}/(4
  \pi)}$~\cite{Cribiori:2022cho}. For $(e'Q) M_*^{(d-2)/2}/(2 M \sqrt{\pi}) \ll 1$, it follows that $c \sim M$, which leads to the
non-extremal relation between $S$ and
$c$, i.e. $c \sim M_* \ S^{(d-3)/(d-2)}$. However, for the near-extremal case
with $M \sim (e'Q) M_*^{(d-2)/2}/\sqrt{4\pi}$, the scaling of the $c$
and of the temperature in terms of $S$ is considered
in~\cite{Basile:2024dqq}, and one has to expand the square root to see that the leading term cancels and the sub-leading term
provides
\begin{equation}
  c \sim M_* \ \sqrt{\beta} \ S^{(d/2-2)/(d-2)} \,,
\end{equation}
  which in turn leads to
\begin{equation}
    T_{ne} \sim M_* \ \sqrt{\beta} \ S^{-d/(2d-4)} = M_* S^{-1/(d-2)}
    \ \sqrt{\beta/S} \, ,
  \end{equation}
  with $M = M_* S^{(d-3)/(d-2)} (1 + 2 S^{-1})$ and
  $e'Q M_*^{(d-2)/2}/\sqrt{4\pi} = M_* S^{(d-3)/(d-2)} [ 1 + (2 +
  \beta) S^{-1}]$, and where $\beta$ is an order-one parameter that controls the differences between the 
masses and charges of particle species and hence also the difference
between mass and charge of the associated near-extremal black hole. Therefore,
\begin{equation}
  T_{ne} = T_s \sqrt{\beta/S}
\label{Tne}
\end{equation}
and so it follows that
\begin{equation}
  \Gamma_{ne} \sim T_{ne} = \sqrt{\beta/S} \ \Gamma_s \, .
\end{equation}
Altogether, the evaporation rate of near-extremal black holes would be
suppressed by a factor of $\sqrt{\beta/S}$ with respect to that of Schwarzschild black holes of the same mass.

Next, in line with our stated plan, we investigate how near-extremal
black holes could modify the PBH range given in Eq.~(\ref{massrange}). To do so, we
consider a black hole with $M \sim 10^5~{\rm g}$. From (\ref{tempes}) we see that
such a black hole has a temperature $T_s \sim 4~{\rm GeV}$. This
means that $c_i(T_s)$ receives a contribution of 2 from photons, 6
from neutrinos, 12 from charged leptons (electrons, muons, and taus), 48 from quarks (up,
down, strange, and charm), and 24 from gluons, yielding  $\sum_{i}
      c_i (T_s) \ \tilde f
      \  \Gamma_s =   45$. Substituting these figures into
      (\ref{cinco}) we find that the lifetime of a $10^5~{\rm g}$
      Schwarzschild black hole is $\tau_s \sim  10^{-5}~{\rm yr}$. For a near extremal black hole of the
      same mass, the temperature would be $T_{ne} \sim 10^{-5} \sqrt{\beta}~{\rm
        eV}$, where we have used (\ref{Tne}). Bearing this
      in mind we find that the lifetime
      of the near-extremal black hole would be $\tau_{ne} \sim 15/\sqrt{\beta}~{\rm
        Gyr}$. Now, the temperature of the near-extremal black hole is
      below the CMB temperature and hence there are no constraints
      from electromagnetic signals. The bound simply comes from the
      black hole survival probability. Then, a rough order of magnitude estimate suggests that if there were 5D primordial near-extremal black holes in nature, then a PBH all-dark-matter
interpretation would be possible in the mass range 
      \begin{equation}
        10^5 \sqrt{\beta} \alt M/{\rm g} \alt 10^{21} \, .
\end{equation}
Note that by tuning the $\beta$ parameter we can have a PBH
all-dark-matter interpretation with very light 5D black holes.  Note
also that
\begin{equation}
  \hat c = c/M \sim \sqrt{\beta/S} \,, 
\end{equation}
which quantifies the near-extremality, is very small because of the large
 entropy.

\section{Conclusions}
\label{sec:6}

We have studied the decay rate of near-extremal black
holes within the context of the dark dimension. Using generic entropy
arguments we have demonstrated that Hawking
evaporation of higher-dimensional near-extremal black holes proceeds
at a slower rate than the corresponding Schwarzschild black holes of
the same mass. Armed with our findings we have shown that if there were 5D primordial
near-extremal black holes in nature, then a PBH all-dark-matter
interpretation would be possible in the mass range  $10^{5}
\sqrt{\beta} \lesssim M/{\rm
  g} \lesssim 10^{21}$, where $\beta$ is a parameter that controls the difference between mass and charge of the associated near-extremal black hole.

The possible existence of near-extremal PBHs evaporating today remains
an open question. We have discussed herein an interesting possibility
in which the black hole is charged under a generic unbroken $U(1)$
symmetry of the dark dimension, whose carriers are always much heavier
than the temperature of the black hole, and so the charge does not get
evaporated away from the black hole and remains therefore
constant. Alternatively, it has been speculated
in~\cite{Arbey:2019jmj} that PBHs may have been formed with a spin
above the 
Thorne's limit $a_* < 0.998$ of astrophysical objects~\cite{Thorne:1974ve}, and actually
near the Kerr extremal value $a_* < 1$ set by the third law of
thermodynamics~\cite{Bardeen:1973gs}, where $a_* = a/M$, with
$a \equiv J_k/M$ the spin parameter and $J_k$ the black hole angular
momentum. If this were the case, then PBHs may be still spinning
today. Further investigation along these lines is obviously important
to be done.

We end with an observation.
 The spectrum of graviton emission from black hole evaporation peaks
 at a frequency which is an order one factor times the temperature of
 a Schwarzschild black hole, $\omega_{\rm peak} \sim T_s$~\cite{Ireland:2023avg}. For
 ultra-light black
 holes $M \sim 10 M_*$, the spectrum peaks at $\omega_{\rm peak} \sim
 M_* (M_*/M)^{1/2} \sim 10^8~{\rm GeV}$. It was recently speculated 
 that in scenarios with large-extra dimensions graviton emission from ultra-light PBHs may be
observed by future gravitational wave detectors~\cite{Ireland:2023zrd}. Here we generalized
the estimate of~\cite{Ireland:2023zrd}  to the dark dimension scenario. Firstly, we note that
after accounting for the redshift in energy density and frequency due
to the cosmological expansion between evaporation and today the
gravitational wave spectrum of a $10 M_*$ PBH would have a peak at a frequency of $10^{12} \alt f/{\rm Hz}
\alt 10^{14}$; see Fig.~4 of Ref.~\cite{Ireland:2023zrd}. This
frequency is in the range of
JURA~\cite{Beacham:2019nyx} and OSQAR II~\cite{OSQAR:2015qdv}
experiments. Secondly, the gravitational wave energy density
can be estimated from Fig.~5 of Ref.~\cite{Ireland:2023zrd} and is given by $10^{-8} <
\Omega_{\rm GW} h^2 < 10^{-6}$. Finally, we note that such a
gravitational wave energy density is orders of magnitude below the
current sensitivity of JURA and OSQAR II~\cite{Ireland:2023avg}.

\section*{Acknowledgements}
The work of L.A.A. is supported by the U.S. National Science
Foundation (NSF Grant PHY-2112527). I.A. is supported by the Second Century Fund (C2F), Chulalongkorn University. The work of D.L. is supported by the Origins
Excellence Cluster and by the German-Israel-Project (DIP) on Holography and the Swampland.

\section*{Appendix}

For completness, in this Appendix we provide a concise summary of the salient
characteristics of 4D RN-dS black holes.
For $d=4$, the blackening function (\ref{uofr}) is given by
\begin{equation}
  U(r) = 1 - \frac{2GM}{r} +
  \frac{(e'Q)^2G}{4\pi \ \ r^{2}} - \frac{r^2}{\ell_4^2} \, .
\label{uofr4}
\end{equation}
With the change of variables ${\cal M} = GM$ and ${\cal Q} = \sqrt{G} Q e'
/\sqrt{4\pi}$ we can rewrite (\ref{uofr4}) in the compact form
\begin{equation}
  U(r) = 1 - \frac{2{\cal M}}{r} + \frac{{\cal Q}^2}{r^2} -
  \frac{r^2}{\ell_4^2} \, .
\label{calU}  
\end{equation}
4D RN-dS configurations generally admit three
horizons, which are located at $r = r_h$ where (\ref{calU}) vanishes,
i.e. $\left. U(r)\right|_{r = r_h} = 0$, yielding a quartic polynomial. The number of real roots is dictated by the sign of the discriminant
locus $D$ of the quartic polynomial
\begin{equation}
\frac{D}{16} = \frac{{\cal M}^2}{\ell_4^2} - \frac{
 {\cal Q}^2}{\ell_4^2} - \frac{27 {\cal M}^4}{\ell_4^4} + \frac{36
 {\cal M}^2 {\cal Q}^2}{\ell_4^4} - \frac{8 {\cal Q}^4}{\ell_4^4} -
\frac{16 Q^6}{\ell_4^6} \, .
\end{equation}
For $D \geq 0$, the quartic polynimial has four real-valued
roots. However, one of them is always negative and therefore
unphysical. Then, the spacetime can have a maximum of three causal
horizons, which are dubbed: the Cauchy (a.k.a. inner) horizon $r_-$, the event
(a.k.a. outer) horizon $r_+$, and the cosmological horizon
$r_c$. Note that $r_+$ and $r_c$ are the horizons which are accessible to an observer outside of the black hole. 

Following~\cite{Romans:1991nq}, we define the phase space of 4D RN-dS black holes as the 3D
parameter space spanned by the mass ${\cal M}$, charge ${\cal Q}$ and
de Sitter radius $\ell_4$. To respect the Cosmic Censorship Conjecture~\cite{Penrose:1969pc},  we require ${\cal M} \geq 0$ and $D \geq 0$, which ensures that all three horizons are real and satisfy $r_-
\leq r_+ \leq r_c$. We also require $\ell_4 \geq 0$ to exclude
AdS. We refer to the region that respects these conditions as physical
phase space $D$. 

\begin{figure}
\postscript{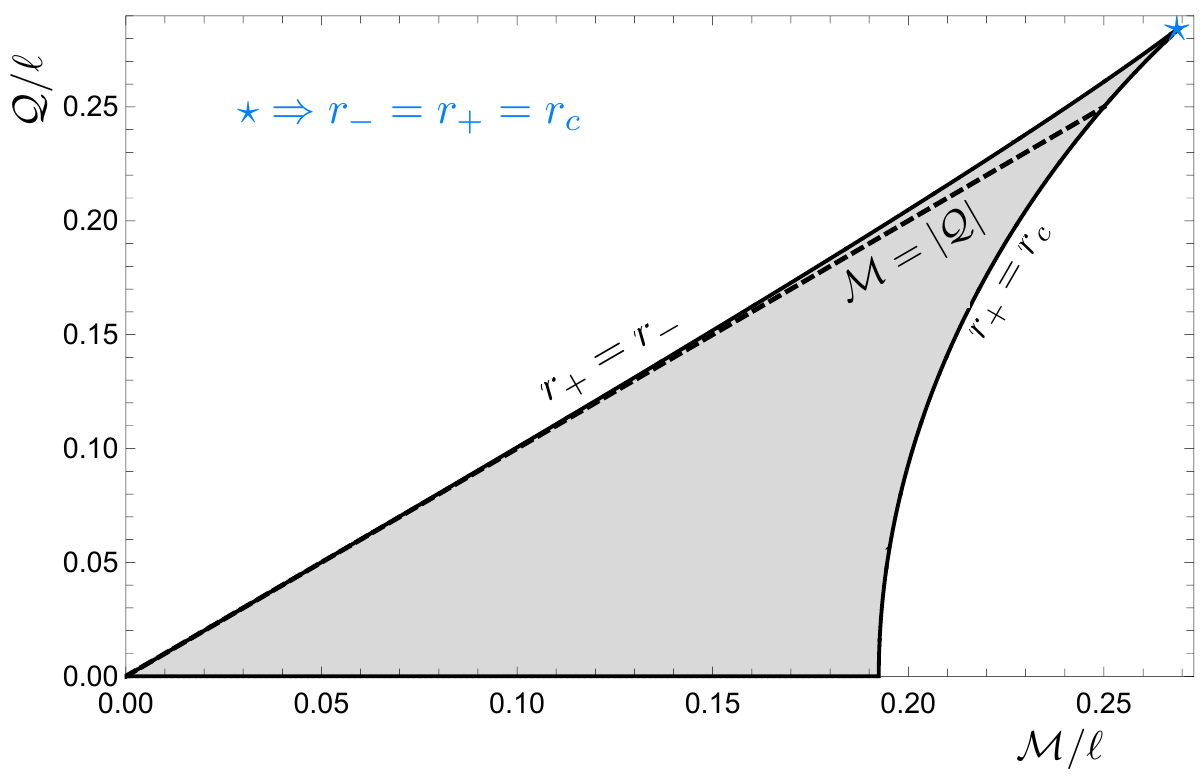}{0.9}
  \caption{The family of 4D RN-dS black holes. The gray shaded region represents the physical phase space of sub-extremal solutions. The boundary of this
    allowed region has two branches: the left (or cold) branch
    corresponds to RN-dS
    extremal black holes and the right branch corresponds to charged Nariai black holes, for which the event and cosmological horizons coincide. The blue star
    where the two branches intersect, stands for the ultracold
    solution. The dashed line indicates the lukewarm solutions with ${\cal
      M} = |{\cal Q}|$, where the Cauchy and event horizons have the
    same temperature. The ${\cal Q} = 0$ axis of neutral black holes
    indicates a big crunch singularity.  Adapted
    from~\cite{Montero:2019ekk}. \label{fig:1}}
\end{figure}

The confluence of two or the three horizons defines an extremal limit
at the boundary of the physical phase space, that we denote by
$\partial D$ and is characterized by $D = 0$. There
are three extremal limits dubbed cold ($r_-=r_+$), Narai ($r_+ = r_c$)
and ultracold ($r_- = r_+ = r_c$). The near horizon geometry for each
of the extremal limits
are AdS$_2 \times S^2$, dS$_2 \times S^2$, and Mink$_2 \times
S^2$, see e.g.~\cite{Castro:2022cuo}. In Fig.~\ref{fig:1} we show the
space of 4D RN-dS solutions. The shaded area is usually referred to as
``shark fin'' due to its shape.

The ultracold near-extremal limit
of the shark fin diagram is a moduli space point that represents Minkowski spacetime
and lies at an infinite distance of any other spacetime independently
of the geodesic path used to reach it~\cite{Luben:2020wix}. The
distance to the ultracold geometry is then consistent with the AdS-DC~\cite{Lust:2019zwm}. On the other hand, the
geometric distance of any spacetime in the 4D RN-dS family to the
origin is finite~\cite{Luben:2020wix}. This implies that black
holes will evaporate back to empty de Sitter space if the FL bound is
satisfied.

On the other hand, in the small curvature limit the dS-WGC implies
that there is at least one state with mass $m$ and charge $q$ satisfying
\begin{equation}
  m^2 < \frac{(e' q)^2}{4 \pi G} - \frac{(e' q)^4}{12 \pi^2 \ell_4^2}
  - \frac{G (e'q)^6}{32 \pi^3 \ell_4^4} + {\cal O} \left(\frac{1}{\ell_4^{6}}
    \right) \,,
\end{equation}
which reproduces the known WGC in flat space
\begin{equation}
e' q > \sqrt{4 \pi G} \ m \left(1 + \frac{G^2 m^2}{2 \ell_4^2} +
  \cdots \right) \,,
\end{equation}
for $\ell_4 \to \infty$~\cite{Antoniadis:2020xso}.

\end{document}